# A new investigation on the characteristics of a W UMa type system LO Andromedae

Peifeng Peng
L3 Advanced Laboratory, Lab Group: Astrolab, Lab Day: Monday, Tuesday, Friday
Submitted: January 11th, 2023, Supervisors: Prof. Mark Swinbank, Prof. Alastair Edge

In this study, a W Ursae Majoris (W UMa) type system LO Andromedae (LO And) was observed over nine nights at Durham to fully investigate its physical properties and update its stellar parameters. After obtaining the observational data in both V-filter and B-filter, the Gaia CCD images of LO And were extracted from the Astrolab internal archive using Linux commands. By applying the "Nightfall" program and local Python scripts, the calibrated light curves of LO And can be plotted, phase-folded, and fitted to derive the period of this system as $0.3804573 \pm 0.0000482$ d. Then, 2D chi-squared heat maps were illustrated to determine LO And's best-fit stellar parameters and estimate their uncertainties, including the mass ratio $q = 0.371 \pm 0.207$, the inclination $i = 78.5 \pm 3.8$ deg, the primary temperature $T_1 = 6290 \pm 358$ K, the secondary temperature $T_2 = 6449 \pm 445$ K, and two fill factors $f_1 = f_2 = 1.02258$. With the help of previous RV surveys and Kepler's third law, the orbital separation $a = 2.62 \pm 0.03$ $R_\odot$, the primary mass $M_1 = 1.22 \pm 0.68$ $M_\odot$, the secondary mass $M_2 = 0.453 \pm 0.357$ $M_\odot$, and the total mass $M_{tot} = 1.67 \pm 0.02$ $M_\odot$ can also be calculated. In addition, the "Nightfall" program built the 3D models and evaluated the spot distribution parameters to visualize the configuration of LO And. There were seven new eclipsing timings added in this study to reconstruct the O-C diagram together with LO And's previous observational data. It was clearly found that the period of LO And is currently undergoing an accelerated increase with a change rate $dP/dt = 2.27 \times 10^{-7}$ d yr$^{-1}$, which was attributed to the mass transfer in this system with a transfer rate $dM_1/dt = 1.43 \times 10^{-7}$ $M_\odot$ yr$^{-1}$ and the light time effect caused by the possible existence of a third star $M_3 = 0.224$ $M_\odot$ orbiting around LO And.

## 1. Introduction

Since the first discovery and observation of binary stars, it has been proved that this specific system is quite common in the main-sequence stars (Jaschek & Gomez 1970). Among their various classifications, eclipsing binary systems mostly attract the interest of astronomers because the magnitude changes can be directly observed from their light curves. One special type of eclipsing binary systems is the W Ursae Majoris (W UMa) type contact binary systems, where the separation is so small that both components fill or overfill their Roche lobes and form a shared atmospheric envelope (Kopal 1955). Studying the W UMa type systems are important to reliably constrain fundamental stellar parameters such as the mass and temperature, and hence shed light on the general understanding of stellar evolution, chemical composition, mass transfer, and formation theory of the W UMa systems and beyond (Paczynski 1971; Yakut & Eggleton 2005).

In this study, LO And with a celestial coordinate (RA: 23h 27m 06.699s; Dec: +45° 33' 22.22") was selected as the target star which was searched from a SIMBAD database of the eclipsing variables catalogue (Malkov et al. 2006), because it can be constantly viewed above 15° altitude to the horizon at Durham every night during the Michaelmas term. Besides, the best observation time was suitably around 22:00 UTC when LO And crossed the local meridian at HA (hour angle) = 0, meaning that the impact of atmospheric disturbance on the observation was safely excluded. LO And ($P \approx 0.3804$ d), as one example of the W UMa type systems, was first discovered by Weber (1963), although it was erroneously classified as a cepheid variable at that time due to the lack of accurate observational equipment. Since then, LO And has been observed over a long history, and all the previous eclipsing timings (times of minima) were collected and studied to evaluate the long-term O-C changes of LO And (Gurol & Muyesseroglu 2005; Nelson & Robb 2015). With seven new eclipsing timings experienced through this study, the O-C diagram can be added with the latest data to keep exploring the reasons of LO And's accelerated period increase. Meanwhile, this study also aims at necessarily updating the period and other stellar parameters of LO And to investigate its physical properties and verify any previous assumptions.

This study was conducted in the Astrolab laboratory at Durham University. Professional training in operating the telescopes was covered to ensure laboratory safety and mitigate any risks prior to formal observations. Ideally, LO And should be observed with the same duration in V-filter and B-filter respectively to better analyze its temperature characteristics. However, due to the bad weather at Durham, successful observations of LO And were only realized over nine clear or partially clear nights, six of which were observed in V-filter and another three in B-filter. As a result, the full light curve of LO And in one period cannot always be obtained over one night, which was perfectly solved by processing local Python scripts to combine different sections of the light curve into one phase. More details about how the local Python scripts worked to extract the uncalibrated light curves from the Gaia images, determine the eclipsing timings of LO And, and plot the phase-folded light curves are further explained in Section 2 of this paper.

In order to achieve the expected outcome, the essential research tool "Nightfall" program (Wichmann 2011) is used in Section 3 of this paper to fit the phase-folded light curves of LO And plotted in Fig. 8. This software can effectively visualize the configuration of LO And using the simulated annealing method to determine the best-fit parameters shown in Table 3 and Table 4, such as the mass ratio, inclination, fill factor, temperature, spot distribution parameters, etc. To better evaluate the correlations between stellar parameters and estimate their uncertainties, 2D heat maps illustrated in Fig. 10 can be plotted from 16 × 16 chi-squared maps sampled in the "Nightfall" program. The 3D models of LO And are also built in Fig. 8 in Section 3.

Section 4 of this paper will discuss the photometric analysis, absolute parameters determination, and the reasons that lead to the period increase of LO And. From the reconstructed O-C diagram in Fig. 11, the quadratic and sinusoidal fit is performed to calculate the parameters demonstrated in Table 6 that describe the orbit of an assumed third star orbiting around LO And. In addition, some statistical relations, such as the *P–M* relations, will also be discussed in Section 4 to further classify whether LO And is a W type or A type W UMa system In Section 5, there will be a short conclusion of this study.

## 2. Methods

### 2.1. *Observations*

Two of four telescopes operated by Durham University Department of Physics were used for the observation of LO And in this study. Draco 2 (2013 SBIG ST9 CCD, 2014-QSI 683ws), a 14" Meade telescope with remote access from the Astrolab laboratory, was primarily used in both V-filter and B-filter over seven nights from October 8$^{th}$ 2022 to November 4$^{th}$ 2022. East-16 (MI CCD), a 16" Meade telescope in the East Dome, was used as an auxiliary equipment to observe LO And in V-filter over two nights of October 24$^{th}$ 2022 and October 28$^{th}$ 2022. For the sake of better image quality, care was taken first to make trial observations and adjust any system parameters for the telescope. Considering that the Earth's atmosphere is at its thinnest directly overhead, the telescopes were steered toward the Zenith to take several images of bright Zenith stars, with the purpose of ensuring that the CCD images were not over-saturated and clear enough after finding the lowest value of full width half maximum (FWHM) at proper telescope focus. More observation details of LO And regarding the filter used, weather conditions, time of observation, and exposure time were all included in the Log of Observations (Table 7) in Appendix A.

### 2.2. *Selecting calibration stars in the Gaia CCD images*

After setting up all the required commands and parameters, the telescope directly connected by the Astrolab computer then started to automatically conduct the observation and save the Gaia CCD images to the Astrolab internal archive. Before extracting the light curves of LO And from the saved Gaia images, it was indispensable to first select two bright calibration stars close to LO And. Since LO And was observed on the ground, the impact of Earth's atmosphere can cause more fluctuations (or noises) in the raw light curves. By analyzing two additional calibration stars marked in Fig. 1, the variation reflected in their light curves should share some similar characteristics to that in the light curve of LO And. Hence, selecting two appropriate calibration stars was a key step to effectively eliminate the impact of Earth's atmosphere, which will also be discussed in Section 2.3 when introducing local Python scripts.

### 2.3. *Extracting the light curves*

To process the Gaia images of LO And, specific Linux directories need to be typed into the terminal to access the data saved in the archive. For instance, all the Gaia images of LO And observed by Draco 2 over the nights of November 4$^{th}$ 2022 can be accessed at the Linux directory: gaia /mnt/archive/draco2/2022/22_11_04. The first Python scripts used in this study is called *aphot.py*, which is an important tool to perform aperture photometry on the CCD images. By creating two files named "cal_sky_position" and "var_sky_position", the celestial coordinates of two calibration stars and LO And as shown in Table 1 were carefully recorded into these two files respectively. The celestial coordinates were determined using the "Pick Object" option in one of the dark-subtracted Gaia images including astrometry. All the Gaia images of LO And taken in one night can then be automatically processed via running *aphot.py* through the Linux command "python /mnt/64bin/auto_astrom/aphot.py" to append all the successfully processed data to a created file called "summary.obs", which contains the Modified Julian Date (MJD) for each data, the uncalibrated magnitudes and their errors of two calibration stars and LO And, and the UTC time for each exposure.

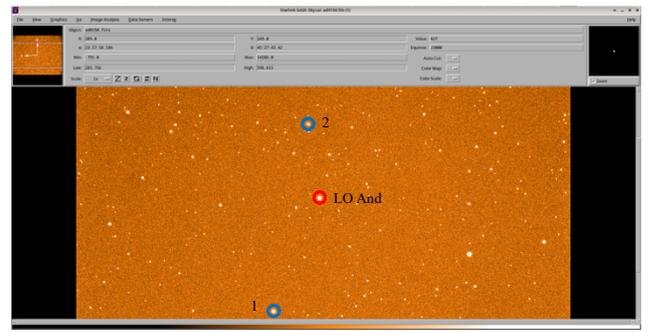

**Fig. 1.** One example Gaia images taken from the telescope's CCD camera. The circle marked in red represents the position of target star LO And, and the other two circles marked in blue represents the position of two calibration stars.

**Table 1.** The celestial coordinates, V magnitude, and B magnitude of two calibration stars and LO And. The magnitudes of two calibration stars were obtained from UCAC4 Catalog in the Gaia images, while the magnitude of LO And was computed from the extracted light curves shown in later sections.

| Parameter | LO And | Calibration Star 1 | Calibration Star 2 |
|---|---|---|---|
| RA (h) | 23:27:06.699 | 23:27:19.351 | 23:27:09.526 |
| Dec (deg) | +45:33:22.22 | +45:28:00.89 | +45:36:53.30 |
| V-mag | 11.292 | 11.417 | 12.167 |
| B-mag | 11.940 | 12.073 | 13.053 |

Finally, to extract the calibrated light curves from the data in the "summary.obs" file, another script called *raw2dif.py3* was used to take into account the impact of Earth's atmosphere on the raw light curves. Theoretically, *raw2dif.py3* was processed with the Linux command "python /mnt/64bin/raw2dif.py3" to assume that the first recorded magnitudes of two calibration stars were correct. Taking these two magnitudes as correction factors (also known as zero-point magnitude) and subtracting the correction factors from the raw light curves, the impact of Earth's atmosphere can be accurately evaluated by working out the offset from all the subsequent observations back to the standard magnitude system. The following two equations describe this process in a mathematical form:

$$m_{cal} = m_{zp} - 2.5 \log_{10} C_{cal}, \quad (1)$$
$$m_{target} = m_{zp} - 2.5 \log_{10} C_{target}, \quad (2)$$

Where $m_{cal}$ and $m_{target}$ are the apparent magnitude of the calibration star and LO And, $m_{zp}$ is the zero-point magnitude, and $C_{cal}$ and $C_{target}$ are the total counts of the calibration star and LO And received on the telescope's CCD camera. If two calibration stars are not variable, the processed light curves of two calibration stars after applying the correction factors to eliminate the atmosphere fluctuations were plotted as two nearly flat lines as seen from the space. As a result, by averaging the difference calculated from two correction factors and applying it to the target star, the calibrated light curve of LO And can be obtained with error bars included, which is expressed in the equations below by combining equation (1) and (2):

$$m_{target} = m_{cal} - 2.5 \log_{10} \frac{C_{target}}{C_{cal}}, \quad (3)$$

Note that to calibrate the magnitude of LO And from the uncalibrated data in the "summary.obs" file, it can be easily realized by directly adding the magnitudes of two calibration stars after the script *raw2dif.py3* and running them together in Linux. Also, the light curves of LO And extracted from Gaia images acquired at different nights should be saved into their corresponding Linux directories in order to avoid mixing their data files. Taking the same example of LO And observed in B-filter over the nights of November 4th 2022, all the data files after running *aphot.py* and *raw2dif.py3* were saved in the following Linux directory: /home/astrolab/EB/LOAnd/22_11_04/draco2/B/, where "EB" represents the abbreviation of eclipsing binary. The entire data analysis process of the script *raw2dif.py3* is illustrated in Fig. 2.

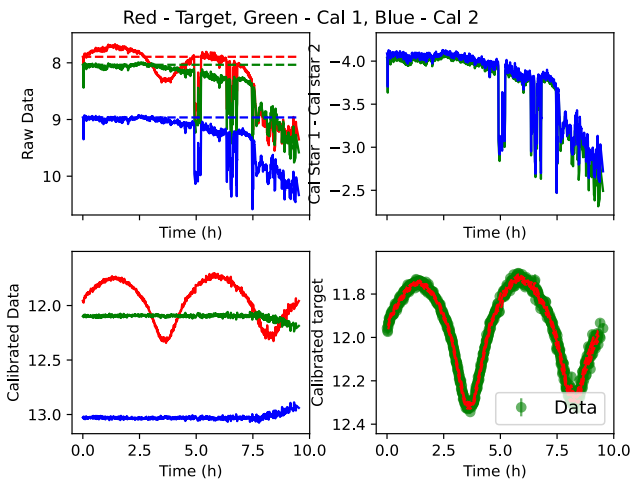

**Fig. 2.** The example light curve analysis by applying the zero-point magnitude calibration when LO And was observed in B-filter over the night of November 4th 2022 using Draco 2.

### 2.4. Aperture Photometry and SNR optimization

When running the Python script *aphot.py*, it automatically performed aperture photometry using the "Results in data counts" option in the Gaia images to measure the total counts and noises of LO And's light falling inside the circular aperture. Basically, there are three circular circles used in aperture photometry, with innermost circle determining the total counts of the target star and outer two circles considering the image noises. The default values of aperture photometry saved in a file called "automag_driver" are the radii of three circles, corresponding to 5, 15, and 25 arcseconds. Although the default values work for most cases, it is still necessary to check these answers to ensure that they also work for LO And. The method used for justification is the signal-to-noise ratio (SNR) optimization, the principle of which is to prove that whether the SNR is optimized at the default radius of inner circle (5 arcseconds) by computing the value of SNR at each inner circle radius. In this study, the sum in aperture (total counts), mean counts (signal), and error in counts (noise) were all recorded in a SNR file, and a graph of the variation of SNR and total counts to aperture radius is then plotted in Fig. 3 by utilizing Matplotlib in Python. Specifically, SNR was calculated as the mean counts divided by error in counts. From the SNR plot in Fig. 3, it clearly suggests that the optimized SNR is reached at an aperture radius of approximately 2.2 arcseconds. However, since the total counts at this aperture radius are underestimated, the default value of inner circle radius in the "automag_driver" file was kept at 5 arcseconds to better process the Gaia images of LO And, even though the SNR is not optimized.

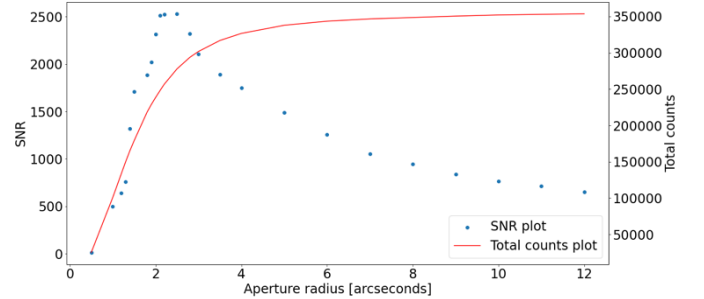

**Fig. 3.** The variation of SNR and total counts to aperture radius plotted in one graph. When the aperture radius is equal to around 2.2 arcseconds, the value of SNR is optimized.

### 2.5. Data reduction techniques and period determination

The photometric precision of the calibrated light curves can be further improved by applying another data reduction technique called flat fielding correction. The images after performing flat fielding correction will be uniformly illuminated by removing any optical variations, such as dust specks on the surface of the telescope's CCD camera, higher sensitivity in the central pixels of CCD, and vignetting effects (Manfroid 1995). The newly developed flat fields for the Michaelmas term of 2022 can be found in the directory: ls /mnt/share/2022_mich_flats.

Another data reduction technique used in this study is called the Heliocentric correction, which is performed to correct the time in any data file from Modified Julian Date (MJD) to Heliocentric Modified Julian Date (HMJD). After processing the Python script *raw2dif.py3*, it will return a simplified version of "summary.obs" data file called "results.diff", which only contains the MJD for each exposure, and the calibrated magnitudes and their errors of LO And. Since the position of LO And is beyond our Solar System, the precision of event times observed on the Earth is slightly hampered due to the relative motion between the Earth and the Sun. To correct this small difference, a Python script called *cor2hjd* was processed with the Linux command "/mnt/64bin/cor2hjd" in this study to convert MJD in the "results.diff" data file to HMJD. Moreover, this procedure will be imperative to accurately measure the eclipsing timings of LO And in Section 2.6.

In addition, clipping the observational data with large errors is also a significant data reduction technique to achieve precise photometry. To plot a error distribution graph, the "results.diff" data file was first uploaded to a graphical viewer called "Topcat". Then, a histogram of data spanned by their errors can be plotted in "Topcat". If the observation is very accurate, the histogram should be spanned around lower values of error with a shape of Gaussian distribution. However, it turned out that there were always some data with large errors deflected to the main distribution. By running a self-written Python script, these data were manually clipped from the calibrated data file to reduce the uncertainty. Thus, the calibrated light curves will be smoother with smaller errors for each data.

running a Python script called *fastsolve.py* with a basic theory of phase dispersion minimization. First of all, select a guess value for the period of LO And, and divide the HMJD in all the "results.diff" data file by the selected period and subtract the integer number of this selected period, so that all the fractional light curves can be folded together in one phase. After this, bin the light curves into several bunches of chunks and measure the dispersion between period intervals. For each selected period, take the average of all the dispersions calculated from each bunch of data and plot a graph of dispersion as a function of the period. If the selected period is close to the correct answer, the averaged dispersion should be very low, which corresponds to a local minimum. In this study, by selecting the period of LO And as 0.38 d based on previous papers (Gurol & Muyesseroglu 2005; Nelson & Robb 2015) and a binned number of 100, the Python script *fastsolve.py* was run through the Linux command "python /mnt/64bin/auto_astrom/fastsolve.py" to automatically process all the data files and combine them to plot the phase-folded light curves in Fig. 4. The initial result of the period was computed as 0.3804 d in Fig. 5.

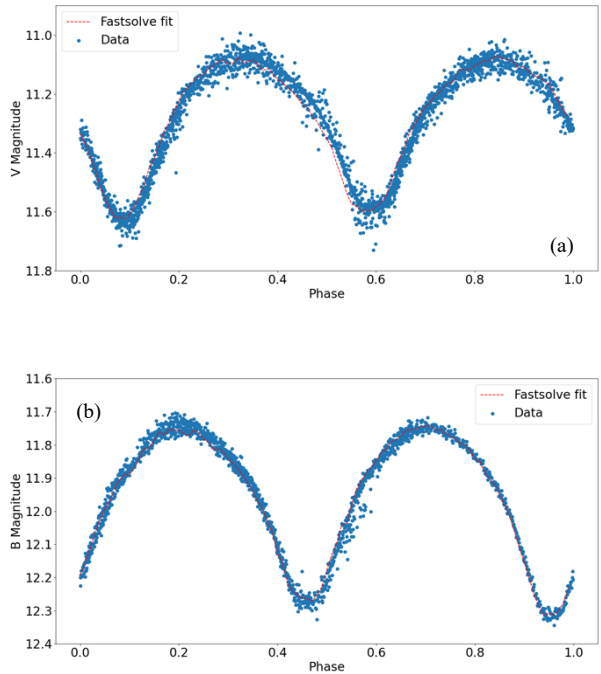

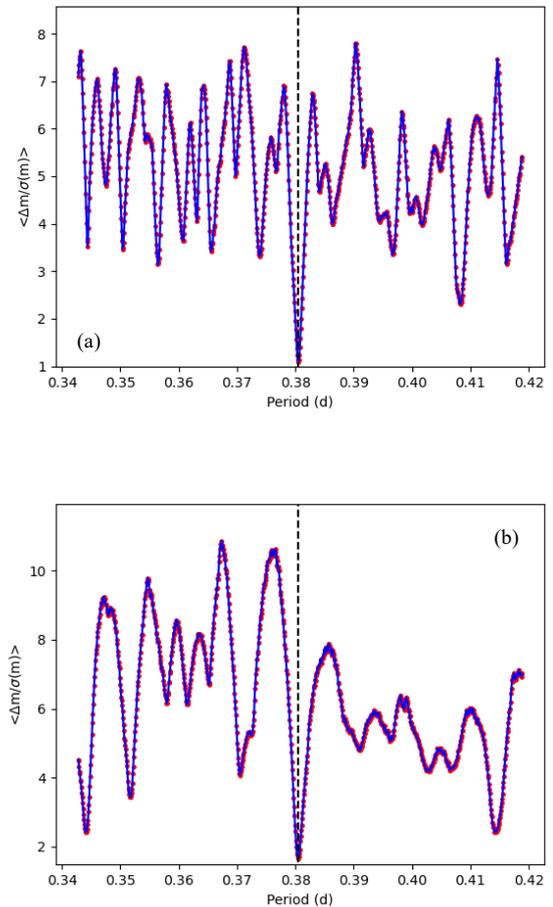

**Fig. 4.** The phase-folded light curves of LO And obtained from running *fastsolve.py*, where panel (a) is for V-filter and panel (b) is for B-filter. The fastsolve fit represents the averaged position of all data points at each phase. By taking the average of V-filter data and B-filter data, the observed V-magnitude and B-magnitude can be computed as 11.292 and 11.940 respectively shown in Table 1.

When it comes to period determination, the fractional light curves of LO And obtained from different nights should be combined to plot a complete, phase-folded light curve. In this study, this process can be perfectly accomplished by

**Fig. 5.** The initial result of period determination using phase dispersion minimization after phase folding the light curves of LO And, where panel (a) is for V-filter and panel (b) is for B-filter. The period is determined to be 0.3804 d from both plots.

Although the initial period seems to be satisfying, it is far from a precise result for which the period needs to be reported in more decimal places with its uncertainty estimated. Therefore, the script *fastsolve.py* was manually modified to measure the dispersion within a much narrower range of period around 0.3804 d and change the binned number to 200. For the data in V-filter, the searching period was ranged from 0.38044 d to 0.38045 d with a result of 0.3804406 d. While for the data in B-filter, the searching period was ranged from 0.38045 d to 0.38055 d, with a result of 0.380505 d. Since direct estimations of the period and its uncertainty were difficult, a statistical method for data reduction called "Jackknife" (Miller 1964) was performed to clip a dataset observed over one night, recalculate the period using *fastsolve.py*, repeat the above procedure, and then take the average of all the recalculated periods. The period of LO And obtained from the V-filter data was computed as $0.3804384 \pm 0.0000042$ d and the period of LO And obtained from the B-filter data was computed as $0.38047625 \pm 0.00009636$ d, where the uncertainties of these two periods were estimated by analyzing the standard deviations of all the recalculated periods after performing "Jackknife". The final period of LO And was thus determined as $0.3804573 \pm 0.0000482$ d.

### 2.6. *Eclipsing timing measurements*

In this study, seven new eclipsing timings (four primary minima, three secondary minima) of LO And were experienced and carefully measured by fitting the local minima with the polynomial fit (polyfit) in Python. To fit the local minima more accurately, only a small fraction of light curves around each local minimum was selected to apply the polynomial fit with a degree of 4. One example fit is plotted in Fig. 6 together with its normalized residuals, while the rest of the eclipsing timings are determined in Fig. 12 in Appendix B. To test whether the polynomial fit in Fig. 6 is a good fit, a chi-squared hypothesis test was performed to calculate a reduced chi-squared of 0.766 and a P-value of 0.962, implying that the polynomial fit is relatively reasonable to fit the data. Besides, all the measurements of eclipsing timings are demonstrated in Table 2.

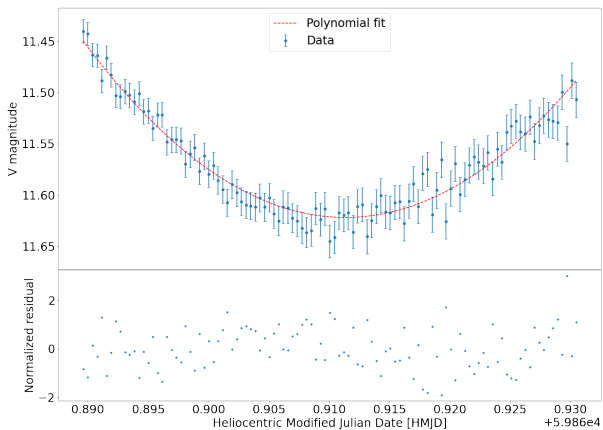

**Fig. 6.** The fractional light curves of LO And observed in V-filter over the night of October 8$^{th}$ 2022 using Draco 2. The error bars are also included in this plot. Since no obvious trend is revealed in the normalized residual plot, it indicates that the data are consistent well with the polynomial fit.

**Table 2.** The eclipsing timing measurements of LO And.

| HMJD | Minima Type | Filter Type |
|---|---|---|
| 59860.91121 | I | V |
| 59866.99892 | I | V |
| 59870.04264 | I | V |
| 59880.88478 | II | V |
| 59880.88536 | II | B |
| 59887.92255 | II | B |
| 59888.11577 | I | B |

## 3. Results

The phase-folded light curves of LO And in V-filter and B-filter were both fitted by the "Nightfall" program, an astrophysical public research tool (Wichmann 2011) to fit the observational data of eclipsing binary systems and return the best-fit stellar parameters, including the mass ratio $q$, inclination $i$, two fill factors $f_1$ and $f_2$, primary and secondary temperature $T_1$ and $T_2$, and spot distribution parameters. In this study, two binned data files of LO And in V-filter and B-filter collected after running *fastsolve.py* were used as the input data files for the "Nightfall" program. Then, these two binned data files were fitted separately as shown in Fig. 9, and the best-fit stellar parameters were determined by utilizing the simulated annealing method provided in the "Nightfall" program. The best-fit stellar parameters in two filters and the final determination of stellar parameters are listed in Table 3, while the spot configurations are included in Table 4. The 3D models included in Fig. 9 can also be built from the "Nightfall" program to better visualize the configuration of LO And.

Another statistical method offered in the "Nightfall" program was the chi-squared mapping, which can be applied to estimate the uncertainties of the best-fit stellar parameters. Based on the parameter configuration of LO And obtained from simulated annealing, while holding all other parameters constant, only two stellar parameters were changed each time when processing the chi-squared maps. After acquiring the output of $16 \times 16$ chi-squared maps, the 2D chi-squared heat maps can be further plotted using Matplotlib in Python to evaluate the correlations between stellar parameters. In this study, by sampling the chi-squared maps from the V-filter and B-filter stellar parameters of LO And, the mass ratio $q$ to the inclination $i$, fill factor 1 to fill factor 2, and primary temperature to secondary temperature 2D heat maps were all illustrated in Fig. 10 with chi-squared contours included.

## 4. Discussion

### 4.1. *Photometric analysis*

From the phase-folded V-filter and B-filter light curves of LO And extracted in Fig. 4, one can further carry out the colour investigation of this eclipsing binary system. To begin with, two binned data files (one for V-filter and the other for B-filter) obtained after running the Python script *fastsolve.py* were used to present the B-V plot of LO And in Fig. 7. Note that the phase of V-filter and B-filter light curves must be aligned by necessary adjustments to accurately calculate the value of B-V. There is no obvious trend revealed in the B-V plot in Fig. 7, which infers that the temperatures of two stars in this binary system are nearly the same due to the direct relationship of B-V and stellar temperature shown in the Hertzsprung-Russell diagram for the main-sequence stars.

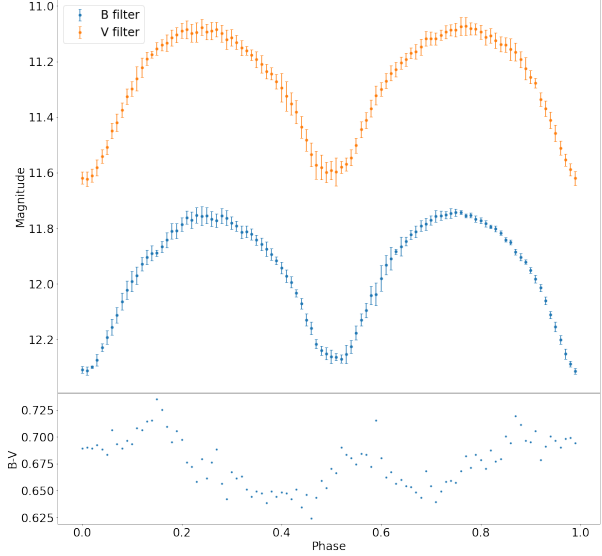

**Fig. 7.** The phase-folded, binned V-filter and B-filter light curves of LO And as well as the B-V plot included underneath, where each data point in the light curve represents the averaged position of other phase-folded data points at that phase. The mean value of B-V is computed as approximately 0.6760 ± 0.0234 mag, where the uncertainty is analyzed by the standard deviation of B-V.

Meanwhile, the colour characteristics of LO And can also be studied by calculating its intrinsic colour index $(B-V)_0$. As a pioneer in first successfully fitting the period-colour diagram of the observed W UMa type systems, Wang (1994) derived the equation below:

$$(B-V)_0 = 0.077 - 1.003 \log_{10} P(\text{day}), \quad (4)$$

Where $P$ is the period of the W UMa type systems. By substituting the period of LO And determined in this study, the intrinsic colour index of LO And is calculated as 0.497953 ± 0.000055 mag, which is inconsistent with the observed value of B-V. This can be explained by a common phenomenon in astronomy known as the interstellar reddening (Trumpler 1930), which arises from the extinction effect when the transmitted light is absorbed or scattered by galactic dust or other matter in the interstellar medium (ISM). Different from redshift, interstellar reddening does not distort the received spectra, meaning that other physical properties of LO And explored by fitting light curves will not be affected. Normally, this effect could be safely ignored after reducing the atmospheric extinction if the observed object is close to our Earth and the object is about to cross the local meridian. However, if the observed object comes from distant galactic sources, interstellar reddening caused by extinction will significantly affect the observed value of B-V. Hence, the colour excess $E(B-V)$ is introduced to better perform the photometric analysis with a necessary calibration process (Eker et al. 2020):

$$E(B-V) = (B-V) - (B-V)_0, \quad (5)$$

Where $(B-V)$ is the observed value of B-V. If the dust map presented and constructed by Schlegel et al. (1998) is applied in this study, the galactic extinction of LO And generated from the dust in the Milky Way disc can be estimated by assuming a standard reddening law (Schlegel et al. 1998; Schlafly & Finkbeiner 2011) to measure the colour excess $E(B-V)$ with the following model:

$$R_V = \frac{A_V}{E(B-V)}, \quad (6)$$

Where $A_V$ is the interstellar extinction in V-filter, and $R_V$ is the total-to-selective extinction parameter with a mean value of 3.1 in the diffuse ISM (Schlegel et al. 1998). For LO And, the value of $A_V$ is determined from the NASA/IPAC Galactic Dust Reddening and Extinction map. According to the Schlegel et al. (1998, SFD) dust map, the galactic extinction is estimated as $A_{V,\text{SFD}} = 0.5353$ mag, while Schlafly & Finkbeiner (2011, S & F) improved the accuracy of the standard reddening law with a 14% recalibration of the SFD dust map by taking into account the blue tip of the stellar locus to give their new estimate as $A_{V,\text{S \& F}} = 0.4604$ mag. As a result, one can obtain two values of $E(B-V)$ for LO And, with one previous estimate of $E(B-V)_{\text{SFD}} = 0.1769 \pm 0.0077$ mag and the other estimate of $E(B-V)_{\text{S \& F}} = 0.1522 \pm 0.0066$ mag. The intrinsic colour index of LO And can therefore be recalculated as $(B-V)_{0,\text{SFD}} = 0.4991 \pm 0.0246$ mag and $(B-V)_{0,\text{S \& F}} = 0.5238 \pm 0.0243$ mag, where the uncertainties are ascertained from error propagation.

Since the S & F estimation for LO And is more accurate compared with the SFD estimation, after subtracting the colour excess $E(B-V)$ from the observed B-V, it points out that the colour-period relation given by Wang (1994) might require further corrections, even though it agrees well with the SFD estimation. What's more, with reference to the fundamental astrophysical parameters listed in Eker et al. (2020), both components of LO And can be classified as F7, F8, or G0 type main-sequence stars. The temperature of LO And fitted by the "Nightfall" program can also be checked with the following colour-temperature relation derived by Eker et al. (2020) for the stars cooler than 10000 K:

$$\begin{aligned}\log_{10} T_{\text{eff}} &= 0.07569(\pm 0.012) \\ &\times (B-V)_0^2 - 0.38786(\pm 0.01368) \\ &\times (B-V)_0 + 3.96617(\pm 0.00338),\end{aligned} \quad (7)$$

Where $T_{\text{eff}}$ is the effective temperature of LO And. Using the S & F estimation result $(B-V)_{0,\text{S \& F}}$, the effective temperature is calculated as $T_{\text{eff}} = 6078 \pm 180$ K, which is consistent with both the "Nightfall" fitted temperature shown in Table 3 and the temperature $T_{\text{eff}} = 5873.508$ K published by Gaia Data Release 2 (Gaia Collaboration et al. 2016, 2018, Gaia DR2).

Although the "Nightfall" program helped fit most stellar parameters of LO And, there are still some indirect parameters that cannot be obtained from fitting light curves, such as the luminosity. To explore these indirect parameters of LO And, the absolute magnitude in V-filter $M_V$ is first

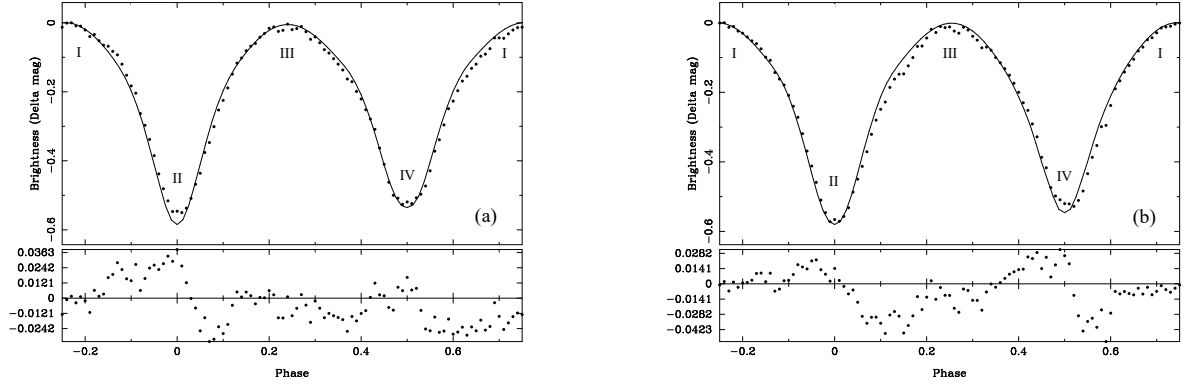

**Fig. 8.** The phase-folded, binned light curves of LO And fitted by the "Nightfall" program, where panel (a) is for V-filter and panel (b) is for B-filter. The residual plot for each filter is also included underneath. After finishing the fitting process, the "Nightfall" program returns a reduced chi-squared value of 0.47587 for the V-filter model, while the reduced chi-squared value for the B-filter model is 0.91142, indicating that both fitting models are reasonable to fit the light curves of LO And.

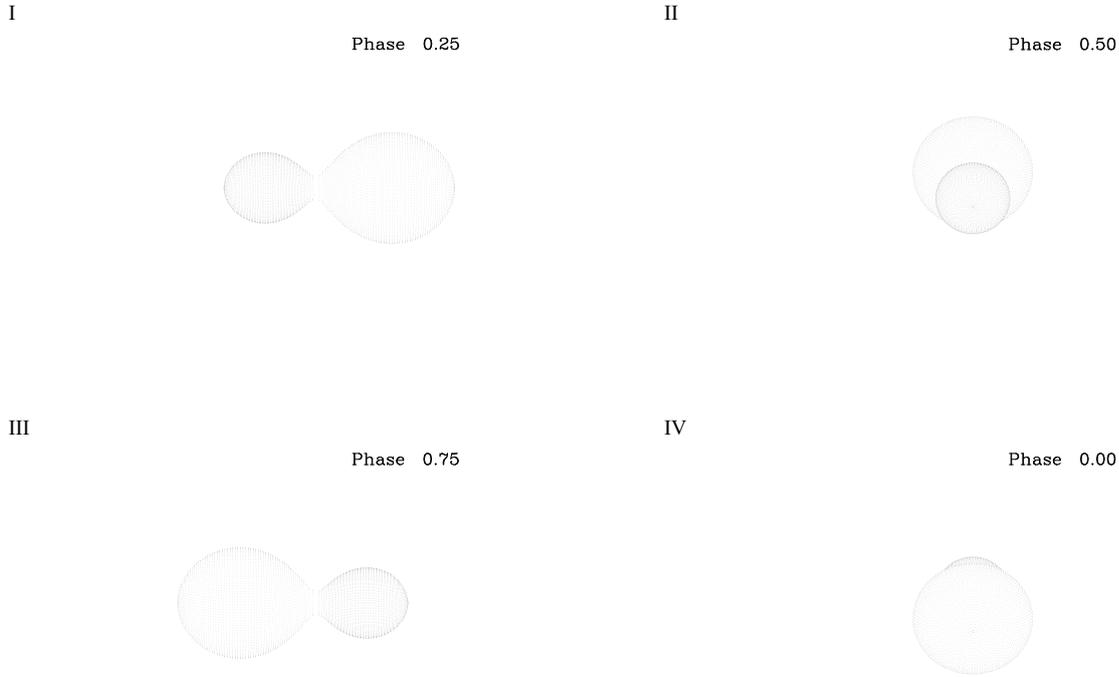

**Fig. 9.** The 3D models of LO And built in the "Nightfall" program. The phase order (I, II, III, IV) in this figure corresponds to the varying trend of light curves labelled in Fig. 8. It can be easily recognized from this figure that both components of LO And overfill their Roche lobes, which is also reflected in the fill factors demonstrated in Table 3 with their best-fit values greater than 1.

estimated by the period-colour relation for contact binary systems as follows (Rucinski & Duerbeck 1997):

$$M_V = -4.44 \log_{10} P(\text{day}) + 3.02(B-V)_0 + 0.12, \quad (8)$$

The absolute V-magnitude of LO And is calculated as $M_V = 3.565 \pm 0.140$ mag using the determined period and intrinsic colour index of LO And. Then, the bolometric magnitude $M_{\text{bol}}$ of LO And can be derived as:

$$M_{\text{bol}} = M_V + \text{BC}, \quad (9)$$

Where BC = 0.033 mag is also determined by Eker et al. (2020). Hence, the bolometric magnitude of LO And is equal to $M_{\text{bol}} = 3.599 \pm 0.140$ mag, which can be quoted to compute the total luminosity of LO And based on the Pogson's relation (Pogson 1856):

$$M_{\text{bol}} - M_{\text{bol},\odot} = -2.5 \log_{10}\left(\frac{L}{L_\odot}\right), \quad (10)$$

Where $M_{\text{bol},\odot} = 4.74$ mag and $L_\odot = 3.828 \times 10^{26}$ W are the bolometric magnitude and luminosity of the Sun taken from the XXIXth IAU General Assembly Resolution B2[1], giving a result of $L = 1.10 \ (\pm 0.14) \times 10^{27}$ W (equivalent to $2.86 \pm 0.37 \ L_\odot$). Meanwhile, the distance of LO And can be computed from:

$$m_V - M_V = 5\log_{10} d - 5 + A_V, \quad (11)$$

[1] https://www.iau.org/static/resolutions/IAU2015_English.pdf

**Table 3.** The stellar parameters of LO And, where the results are determined by averaging the V-filter and B-filter parameters. The uncertainties of two best-fit fill factors are not estimated in this table because the "Nightfall" program cannot compute the chi-squared values for fill factors larger than 1, and hence their best-fit values are not illustrated in the 2D chi-squared heat maps in Fig. 10.

| Stellar parameters | V value | V error | B value | B error | Result | Error | Unit |
|---|---|---|---|---|---|---|---|
| Mass ratio ($q$) | 0.373 | 0.272 | 0.3695 | 0.3119 | 0.371 | 0.207 | |
| Inclination ($i$) | 79.3 | 5.2 | 77.6 | 5.5 | 78.5 | 3.8 | deg |
| Fill factor 1 ($f_1$) | 1.02258 | | 1.02258 | | 1.02258 | | |
| Fill factor 2 ($f_2$) | 1.02258 | | 1.02258 | | 1.02258 | | |
| Primary temperature ($T_1$) | 6390 | 552 | 6190 | 454 | 6290 | 358 | K |
| Secondary temperature ($T_2$) | 6487 | 651 | 6410 | 608 | 6449 | 445 | K |
| Period ($P$) | 0.3804384 | 0.0000042 | 0.38047625 | 0.00009636 | 0.3804573 | 0.0000482 | day |

**Table 4.** The spot configuration of LO And determined by averaging the V-filter and B-filter spot parameters. There is no uncertainty estimated in this table because the "Nightfall" program cannot draw the chi-squared maps for the spot parameters.

| Spot parameters | V spot configuration | B spot configuration | Spot configuration | Unit |
|---|---|---|---|---|
| Primary spot longitude | 308 | 359 | 334 | deg |
| Primary spot latitude | 17.3 | 10.9 | 14.1 | deg |
| Primary spot radius | 6.56 | 6.56 | 6.56 | deg |
| Primary spot dimfactor | 0.504 | 0.504 | 0.504 | |
| Secondary spot longitude | 347 | 304 | 325 | deg |
| Secondary spot latitude | 24.0 | 55.9 | 40.0 | deg |
| Secondary spot radius | 6.87 | 6.87 | 6.87 | deg |
| Secondary spot dimfactor | 0.543 | 0.748 | 0.6455 | |

Where $m_V = 11.292$ is the apparent V-magnitude of LO And. The obtained distance is $d = 284.0 \pm 18.3$ pc, which shows remarkable agreement with the distance $d = 287.9678$ pc on Gaia DR2 (Gaia Collaboration et al. 2016, 2018).

### 4.2. Absolute parameters determination

Apart from the stellar parameters of LO And, the absolute parameters for each component of LO And can also be determined to better investigate the physical properties of LO And. In this study, the primary component of LO And is defined as the more massive and greater component of LO And. From the radial velocity (RV) survey conducted by Nelson & Robb (2015), they performed their self-developed fitting functions with data reduction analysis included to fit their collected observational data and determine the radial velocity as $v_1 = 76.4 \pm 2.8$ km s$^{-1}$, $v_2 = 272.4 \pm 3.3$ km s$^{-1}$. The orbital semi-major axes of two stars can then be calculated as $a_1 = 0.575 \pm 0.021 \ R_\odot$ and $a_2 = 2.05 \pm 0.02 \ R_\odot$ using the following formula:

$$v_{1,2} = \frac{2\pi a_{1,2}}{P}, \quad (12)$$

Where the solar radius is assumed as $R_\odot = 6.957 \times 10^8$ m when converting the unit to report the result of orbital semi-major axis. Under the simple estimation $a = a_1 + a_2$, the separation between two stars is found to be $a = 2.62 \pm 0.03 \ R_\odot$. After this, the total mass of LO And $M_{tot}$ is solved according to the generalized form of Kepler's third law:

$$P^2 = \frac{4\pi^2 a^3}{GM_{tot}}, \quad (13)$$

Which gives a result of $M_{tot} = 1.67 \pm 0.02 \ M_\odot$ by inserting all the known values into equation (13). Since the mass ratio $q = M_2/M_1$ has already been determined in Table 3, the mass for each component can be easily computed as $M_1 = 1.22 \pm 0.68 \ M_\odot$ and $M_2 = 0.453 \pm 0.357 \ M_\odot$. The accuracy of mass calculations is validated by the P–M relations updated by Poro et al. (2021) using the MCMC approach:

$$M_1 = (2.924 \pm 0.075)P + (0.147 \pm 0.029), \quad (14)$$

$$M_2 = (0.541 \pm 0.092)P + (0.294 \pm 0.034), \quad (15)$$

$$M_{tot} = (4.421 \pm 0.273)P + (0.138 \pm 0.009), \quad (16)$$

The P–M relations from equation (14) to (16) estimate a primary mass $M_1 = 1.26 \pm 0.04 \ M_\odot$, a secondary mass $M_2 = 0.500 \pm 0.049 \ M_\odot$, and a total mass $M_{tot} = 1.82 \pm 0.14 \ M_\odot$, indicating that the mass calculations in this study are consistent with the estimated values. Besides, Qian's (2003) statistical investigation showed that LO And should be classified as a W type W UMa system because the derived primary mass is less than the critical mass 1.35 $M_\odot$.

To figure out the radii of both components, their mean fractional radii and orbital semi-major axes are required:

$$a_{1,2} = \frac{R_{1,2}}{r_{1,2}}, \quad (17)$$

Where $r_1 = 0.496 \pm 0.002$ and $r_2 = 0.309 \pm 0.005$ can be calculated by the relation given as (Poro et al. 2021):

$$r = (r_{pole} \times r_{side} \times r_{back})^{\frac{1}{3}}, \quad (18)$$

In this study, the values of $r_{pole}$, $r_{side}$, and $r_{back}$ were obtained by averaging the previous results (Gurol & Muyesseroglu 2005; Nelson & Robb 2015) as they cannot be determined from fitting the light curves of LO And using the "Nightfall" program. As a result, one can obtain the radii as $R_1 = 1.30 \pm 0.02 \ R_\odot$ and $R_2 = 0.812 \pm 0.016 \ R_\odot$. If both components of LO And are assumed to be ideal blackbody radiation

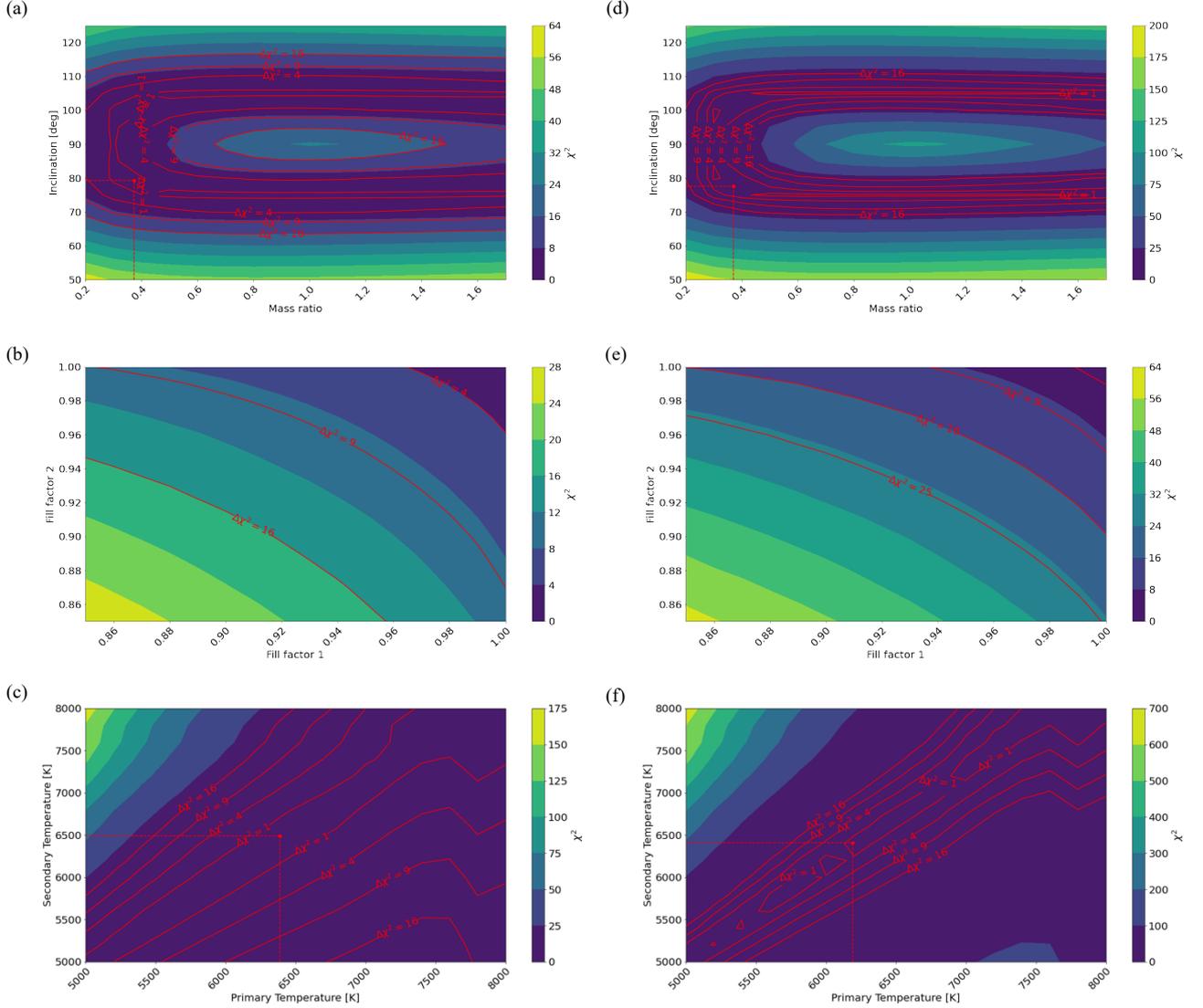

**Fig. 10.** The illustrated 2D heat maps to evaluate the correlations between mass ratio and inclination, two fill factors, and two temperatures of LO And. The uncertainties are estimated by the extremum of $\Delta\chi^2 = 1$ contour. Panel (a), (b), and (c) in the first column are the heat maps for the V-filter parameters, while panel (d), (e), and (f) in the second column are the heat maps for the B-filter parameters.

sources, the Stefan-Boltzmann law can be applied to calculate the stellar luminosity separately:

$$L_{1,2} = 4\pi R_{1,2}^2 \sigma T_{1,2}^4, \quad (19)$$

Where $\sigma = 5.67 \times 10^{-8}$ W m$^{-2}$ K$^{-4}$ is the Stefan-Boltzmann constant. By applying equation (19), the luminosity for each component is computed as $L_1 = 2.39 \pm 0.55\ L_\odot$ and $L_2 = 1.03 \pm 0.29\ L_\odot$. Finally, the absolute magnitude of each component of LO And can be determined with the relation:

$$M_{V1,2} - M_V = -2.5\log_{10}\left(\frac{L_{1,2}}{L}\right), \quad (20)$$

The calculated absolute magnitude is $M_{V,1} = 3.762 \pm 0.318$ mag and $M_{V,2} = 4.679 \pm 0.362$ mag. Meanwhile, the surface gravity $\log(g_1)$ and $\log(g_2)$ were also fitted in the "Nightfall" program. All the derived absolute parameters of LO And are therefore presented in Table 5. One may notice that some other absolute parameters of LO And, such as the limb darkening coefficients and the gravity darkening coefficients (Wilson & Devinney 1971), were not determined in this study, which is due to the lack of information for LO And.

### 4.3. The reconstructed O-C diagram

As seven new eclipsing timings of LO And were experienced during the observation and were carefully determined in Section 2.6, the O-C diagram of LO And can be reconstructed by adding the latest data to update the previous fit and keep studying the period variation of LO And. This section is mainly based on the work of Gurol & Muyesseroglu (2005) and Nelson & Robb (2015), who collected all the previous data of LO And including 15 photographic (pg), 164 visual (vis), 28 photoelectric (pe), and 79 CCD eclipsing timings observed. The details of these data were recorded in the appendix of Gurol & Muyesseroglu (2005) and in Bob Nelson's Database of Eclipsing Binary O-C Files[2]. What's more, Nelson & Robb (2015) further performed their data reduction techniques to clip any observational data with large errors.

To begin with, the following linear ephemeris is used to express the calculated eclipsing timings:

$$\text{HMJD}(\text{MinI}) = t_0 + P_0 \times E, \quad (21)$$

[2] https://www.aavso.org/bob-nelsons-o-c-files

**Table 5.** The absolute parameters of LO And.

| Absolute parameters | Result ± Error | Unit |
|---|---|---|
| Primary mass ($M_1$) | 1.22 ± 0.68 | $M_\odot$ |
| Secondary mass ($M_2$) | 0.453 ± 0.357 | $M_\odot$ |
| Total mass ($M_{tot}$) | 1.67 ± 0.02 | $M_\odot$ |
| Primary radius ($R_1$) | 1.30 ± 0.02 | $R_\odot$ |
| Secondary radius ($R_2$) | 0.812 ± 0.016 | $R_\odot$ |
| Separation ($a$) | 2.62 ± 0.03 | $R_\odot$ |
| Primary V-magnitude ($M_{V,1}$) | 3.762 ± 0.318 | mag |
| Secondary V-magnitude ($M_{V,2}$) | 4.679 ± 0.362 | mag |
| V-magnitude ($M_V$) | 3.565 ± 0.140 | mag |
| Intrinsic colour index $(B-V)_0$ | 0.5238 ± 0.0243 | mag |
| Primary luminosity ($L_1$) | 2.39 ± 0.55 | $L_\odot$ |
| Secondary luminosity ($L_2$) | 1.03 ± 0.29 | $L_\odot$ |
| Luminosity ($L$) | 2.86 ± 0.37 | $L_\odot$ |
| Distance ($d$) | 284.0 ± 18.3 | pc |
| Primary fractional radius ($r_1$) | 0.496 ± 0.002 | |
| Secondary fractional radius ($r_2$) | 0.309 ± 0.005 | |
| Surface gravity $\log(g_1)$ | 4.37 ± 0.04 | |
| Surface gravity $\log(g_2)$ | 4.26 ± 0.02 | |

Where $t_0 = 45071.059$ HMJD is the starting epoch, $P_0 = 0.380435536$ d is the initial period, and $E$ is the eclipsing timings expressed in epochs. Both the values of $t_0$ and $P_0$ are derived by Gurol & Muyesseroglu (2005). After subtracting HMJD(MinI) from the observed eclipsing timings, the $(O–C)$ values can be computed in the upper panel of Fig. 11 and fitted by a quadratic ephemeris:

$$\text{HMJD(MinI)} = c_0 + c_1 \times E + c_2 \times E^2, \quad (22)$$

Where $c_0$, $c_1$, and $c_2$ are the coefficients determined using a weighted least-squares (WLS) method (Kiers 1997). In this study, the weights of photographic (pg) data were set to 0.5, the weights of visual (vis) data were set to 0.1, and the weights of the rest of the data were all set to 1. The fitted values of these three coefficients are listed in Table 6.

Since Gurol & Muyesseroglu (2005) have already proved that only the quadratic fit was not good enough to fit the data because of the sinusoidal varying trend which was attributed to the motion of an assumed third star orbiting around LO And, the quadratic fit is not plotted in this paper. Instead, a more accurate quadratic fit combined with a sinusoidal term $\Delta t$ is plotted to calculate the $(O–C)$ values in the lower panel of Fig. 11. This can be further tested by comparing the weighted sum of the $(O–C)^2$ residuals. For the quadratic and sinusoidal fit, the weighted sum was computed as 0.0289, which was slightly worse than the weighted sum of 0.0258 for the quadratic fit only. This was primarily because the quadratic and sinusoidal fit behaved poorly in modelling the new eclipsing timings, even though the previous eclipsing timings were modelled well by the new fit. The expression for this new ephemeris including the sinusoidal term is performed below (Erdem 2002):

$$\text{HMJD(MinI)} = (t_0 + c_0) \\ + (c_1 + P_0) \times E + c_2 \times E^2 + \Delta t, \quad (23)$$

Where $\Delta t$ comes from the light time effect caused by the elliptical orbit of the third star, which can be expressed as (Irwin 1952; Mayer 1990; Gies et al. 2015):

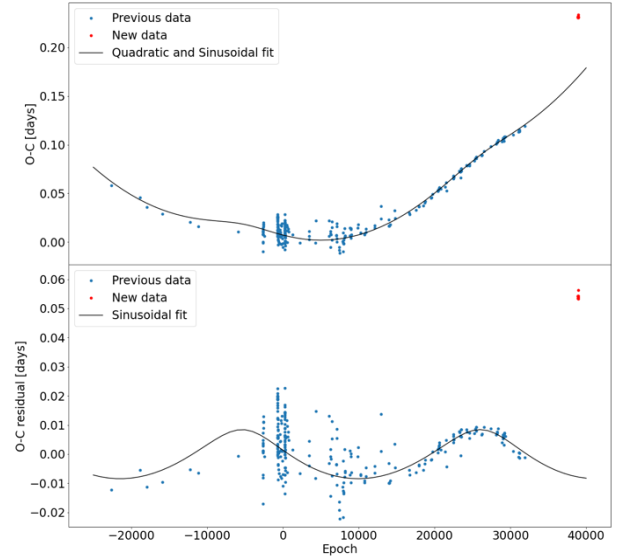

**Fig. 11.** The O-C diagram of LO And, where the quadratic and sinusoidal fit for the upper panel is plotted based on equation (23), and the sinusoidal fit for the lower panel is just the model from equation (24). The data points marked in red represent seven new eclipsing timings collected in this study.

$$\Delta t = \frac{A}{\sqrt{1-e^2\cos^2\omega}} \\ \times \left[\frac{1-e^2}{1+e\cos\nu}\sin(\nu+\omega) + e\sin\omega\right], \quad (24)$$

And $A$ is the semi-amplitude of the light time effect:

$$A = \frac{a_{12}\sin i_3 \sqrt{1-e^2\cos^2\omega}}{c}, \quad (25)$$

In both equation (24) and (25), $a_{12}$ is orbital semi-major axis of LO And to the common center of mass of the triple system, and $e$, $i_3$, $\omega$, and $\nu$ are the parameters describing the orbit of the third star, which represent the eccentricity, inclination, longitude of periastron, and true anomaly respectively. In addition, the mass and period of the third star can be calculated by a mass function (Mayer 1990):

$$f(M_3) = \frac{(M_3\sin i_3)^3}{(M_1+M_2+M_3)^2} \\ = \frac{(a_{12}\sin i_3)^3}{P_3^2}, \quad (26)$$

Where $M_3$ and $P_3$ are the mass and period of the third star. Similarly, by performing the weighted least-squares method, all the third star parameters can be fitted and included in Table 6. If the third star is assumed to be coplanar to the orbit of LO And ($i_3 = i = 78.5 \pm 3.8$ deg), the mass and the period of the third star can be computed as $M_3 = 0.224\ M_\odot$ and $P_3 = 32.7$ yr. Besides, the orbital semi-major axis of the third star $a_3 = 11.2$ AU is resolved after knowing $a_{12} = 1.49$ AU:

$$a_3 = \frac{M_1 + M_2}{M_3} a_{12}, \quad (27)$$

**Table 6.** The parameters used in the fitting models to describe the coplanar orbit of the assumed third star around LO And.

| Parameters | Result | Unit |
|---|---|---|
| Coefficient ($c_0$) | $5.76 \times 10^{-3}$ | |
| Coefficient of $E$ ($c_1$) | $1.83 \times 10^{-7}$ | |
| Coefficient of $E^2$ ($c_2$) | $1.18 \times 10^{-10}$ | |
| Starting epoch ($t_0$) | 45071.059 | HMJD |
| Initial period ($P_0$) | 0.380435536 | day |
| Semi-major axis ($a_{12}$) | 1.49 | AU |
| Semi-major axis ($a_3$) | 11.2 | AU |
| Inclination ($i_3$) | 78.5 | deg |
| Eccentricity ($e$) | 0.305 | |
| Longitude of periastron ($\omega$) | 97.6 | deg |
| Semi-amplitude ($A$) | $8.43 \times 10^{-3}$ | day |
| Period ($P_3$) | 32.7 | year |
| Mass function $f(M_3)$ | $2.92 \times 10^{-3}$ | $M_\odot$ |
| Mass ($M_3$) | 0.224 | $M_\odot$ |
| Period change rate (d$P$/d$t$) | $2.27 \times 10^{-7}$ | d yr$^{-1}$ |
| Mass transfer rate (d$M_1$/d$t$) | $1.43 \times 10^{-7}$ | $M_\odot$ yr$^{-1}$ |

As illustrated in Fig. 11, the period of LO And is undergoing an accelerated increase at present because the current ($O$–$C$) values calculated in this study are larger than the predicted values of the fitting models. In order to estimate the current period change rate of LO And, the equation derived by Nelson & Robb (2015) can be applied:

$$\frac{dP}{dt} = \frac{2c_2}{P_0}, \quad (28)$$

Where $c_2$ is the second order coefficient defined in equation (22). To express the unit of d$P$/d$t$ in d yr$^{-1}$, one need to multiply the number of days in one year (365.24 d) to equation (28), giving a result of d$P$/d$t$ = $2.27 \times 10^{-7}$ d yr$^{-1}$. Furthermore, according to the formula derived by Kwee (1958), the mass transfer rate between two components of LO And is calculated as:

$$\frac{\dot{P}}{P} = 3\left(\frac{M_1}{M_2} - 1\right)\frac{\dot{M}_1}{M_1}, \quad (29)$$

By substituting the period change rate d$P$/d$t$ = $2.27 \times 10^{-7}$ d yr$^{-1}$ into equation (29), a mass transfer rate d$M_1$/d$t$ = $1.43 \times 10^{-7}$ $M_\odot$ yr$^{-1}$ can then be obtained.

For all the third star parameters identified in this study, no uncertainties were estimated because they were found to be extremely large after performing the weighted least-squares method to fit the data, revealing that the parameters demonstrated in Table 6 are less reliable compared with previous studies (Gurol & Muyesseroglu 2005; Nelson & Robb 2015). What's more, considering that the quadratic and sinusoidal fit didn't model the new eclipsing timings well in Fig. 11, other fitting models such as the MPFIT solver (Markwardt 2009) might be a better choice to fit the data and improve the accuracy of the third star parameters.

## 5. Conclusion

In this study, a W UMa type system LO And was observed in V-filter and B-filter over nine nights at Durham using two ground telescopes, Draco 2 and East-16, to reanalyze its various properties by updating the stellar parameters. To first extract the calibrated light curves of LO And from the telescope's Gaia CCD images, several local Python scripts were applied in the Astrolab Linux system to plot and phase-fold the fractional light curves observed at different nights after performing any data reduction techniques at the optimal SNR. Then, the V-filter and B-filter observational data were uploaded to the "Nightfall" program to fit the stellar parameters separately, giving the best-fit results of the mass ratio $q$ = 0.371 ± 0.207, the inclination $i$ = 78.5 ± 3.8 deg, the primary temperature $T_1$ = 6290 ± 358 K, the secondary temperature $T_2$ = 6449 ± 445 K, two fill factor $f_1$ = $f_2$ = 1.02258, and the spot distribution parameters in Table 4. The 2D chi-squared heat maps and the 3D models of LO And were also built in the "Nightfall" program to estimate the uncertainties of the best-fit parameters and visualize the configuration of LO And. Furthermore, the absolute parameters of LO And were analytically determined to explore more detailed physical properties in this contact binary system. After including the latest observational data, the reconstructed O-C diagram clearly suggested an accelerated period increase for LO And. This was perfectly explained by the mass transfer between this system and the light time effect caused by an assumed third star, although the derived third star parameters were less reliable without performing other more accurate fitting models.

The photometric analysis in this study showed that the intrinsic colour index of LO And was $(B - V)_{0, S \& F}$ = 0.5238 ± 0.0243 mag ascertained from the S & F dust map, which corresponded to a spectral type of F7 – G0. Meanwhile, LO And was classified as a W type W UMa system based on the critical primary mass 1.35 $M_\odot$ found by Qian (2003). However, this classification must be verified by further determination because of the large uncertainty estimated for the primary mass $M_1$ arisen from the large uncertainty of the mass ratio $q$ in this study. To reduce the uncertainty of the mass ratio $q$, a $q$-search method (Gurol & Muyesseroglu 2005, Li et al. 2021) could be utilized to evaluate the minimum sum of the weighted squared residuals. Moreover, Binnendijk (1970) stated that for the W type W UMa systems, the temperature of the greater component was cooler, while the temperature of the greater component for the A type W UMa systems was hotter instead. Since the temperature determined for two components of LO And in this study were very close, this also indicated that future research on precise temperature measurements is necessarily required.

*Acknowledgements.* I would like to first thank Prof. Mark Swinbank and Prof. Alastair Edge for their contributions in supervising my Astrolab project and for providing many useful comments and suggestions. This research referred to the public background information of LO And searched in a SIMBAD database of the eclipsing variables catalogue, which is operated at the SAO/NASA Astrophysics Data System Abstract Service. I also wish to acknowledge the public access to many astrophysical research tools which formed the basis of this research, including the "Topcat" graphical viewer developed by Dr. Mark Taylor at the University of Bristol, the "Nightfall" program developed by Dr. Rainer Wichmann at the University of Hamburg, the NASA/IPAC Galactic Dust Reddening and Extinction map at: https://irsa.ipac.caltech.edu/applications/DUST/, and the Gaia Data Release 2 at: https://gea.esac.esa.int/archive/. Thank you to my research group partner Ka-Hunt To for attending our arranged observational sessions on time and for offering technical assistance. All in all, I sincerely appreciate Durham University's Department of Physics for granting Level 3 students the opportunity to undertake their Astrolab research project by utilizing high-tech telescopes in a professional astronomy research laboratory.

# Appendix

## A. Log of Observations for LO And

**Table 7.** The Log of Observations for LO And, including the date and the time of duration the observation took place, the telescope and the filter used, the exposure time, and the notes briefly describing the details and the weather conditions of the observation.

| Date | Time (UTC) | Telescope | Filter | Exposure time (s) | Notes |
|---|---|---|---|---|---|
| 08/10/22 | 20:59:31 – 01:06:39 | Draco 2 | V | 30.0 | Observed by Prof. Mark Swinbank before the start of the course. The night was almost clear with few patchy clouds until midnight. A total number of 380 images were successfully taken by the telescope. FWHM: 3.0 – 5.0 arcseconds. |
| 14/10/22 | 19:40:48 – 00:59:44 | Draco 2 | V | 30.0 | Observed by Ka-Hunt To and me. The night was perfectly clear although it turned cloudy with a few showers after 01:00 am, after which the dome was shut down. A total number of 557 images were successfully taken by the telescope. FWHM: 3.0 – 4.0 arcseconds. |
| 17/10/22 | 20:28:50 – 05:59:55 | Draco 2 | V | 30.0 | Observed by Ka-Hunt To and me. The night was entirely clear till the next morning when the dome was automatically shut down at 06:00 am, although there was a short interval during midnight when the weather became partially cloudy. A total number of 885 images were successfully taken by the telescope. FWHM: 3.0 – 8.0 arcseconds. |
| 22/10/22 | 18:47:36 – 20:22:41 | Draco 2 | V | 30.0 | Observed by Ka-Hunt To and me. The night was only clear for about 2 hours, after which it started to rain lightly. The dome was shut down after 20:24 pm. A total number of 115 images were successfully taken by the telescope. FWHM: 3.0 – 4.0 arcseconds. |
| 24/10/22 | 19:30:52 – 21:03:10 | East-16 | V | 30.0 | Observed by Ka-Hunt To and me. The night was only clear for about 1.5 hours, after which the night was covered by thick clouds which hindered the observation. The dome was shut down after 21:06 pm. A total number of 75 images were successfully taken by the telescope. FWHM: 2.0 – 3.0 arcseconds. |
| 26/10/22 | 19:15:51 – 23:11:00 | Draco 2 | B | 30.0 | Observed by Ka-Hunt To and me. The night was partially clear until midnight when the weather became completely cloudy. A total number of 446 images were successfully taken by the telescope. FWHM: 4.0 – 5.0 arcseconds. |
| 28/10/22 | East-16: 18:38:09 – 23:45:37<br>Draco 2: 19:26:52 – 23:23:59 | East-16 & Draco 2 | V & B | 30.0 | Observed by Ka-Hunt To and me. The V-filter and B-filter observations were conducted simultaneously. The night was partially clear until midnight when the weather became completely cloudy. Both domes were shut down after 23:48 pm. A total number of 352 V-filter images and a total number of 384 B-filter images were successfully taken by two telescopes. FWHM: 2.0 – 5.0 arcseconds. |
| 04/11/22 | 18:27:09 – 03:57:30 | Draco 2 | B | 30.0 | Observed by Ka-Hunt To and me. The night was perfectly clear until 04:00 am when there was a light shower, after which the dome was shut down. A total number of 933 images were successfully taken by the telescope. FWHM: 3.0 – 6.0 arcseconds. |

B. Additional eclipsing timing measurements of LO And

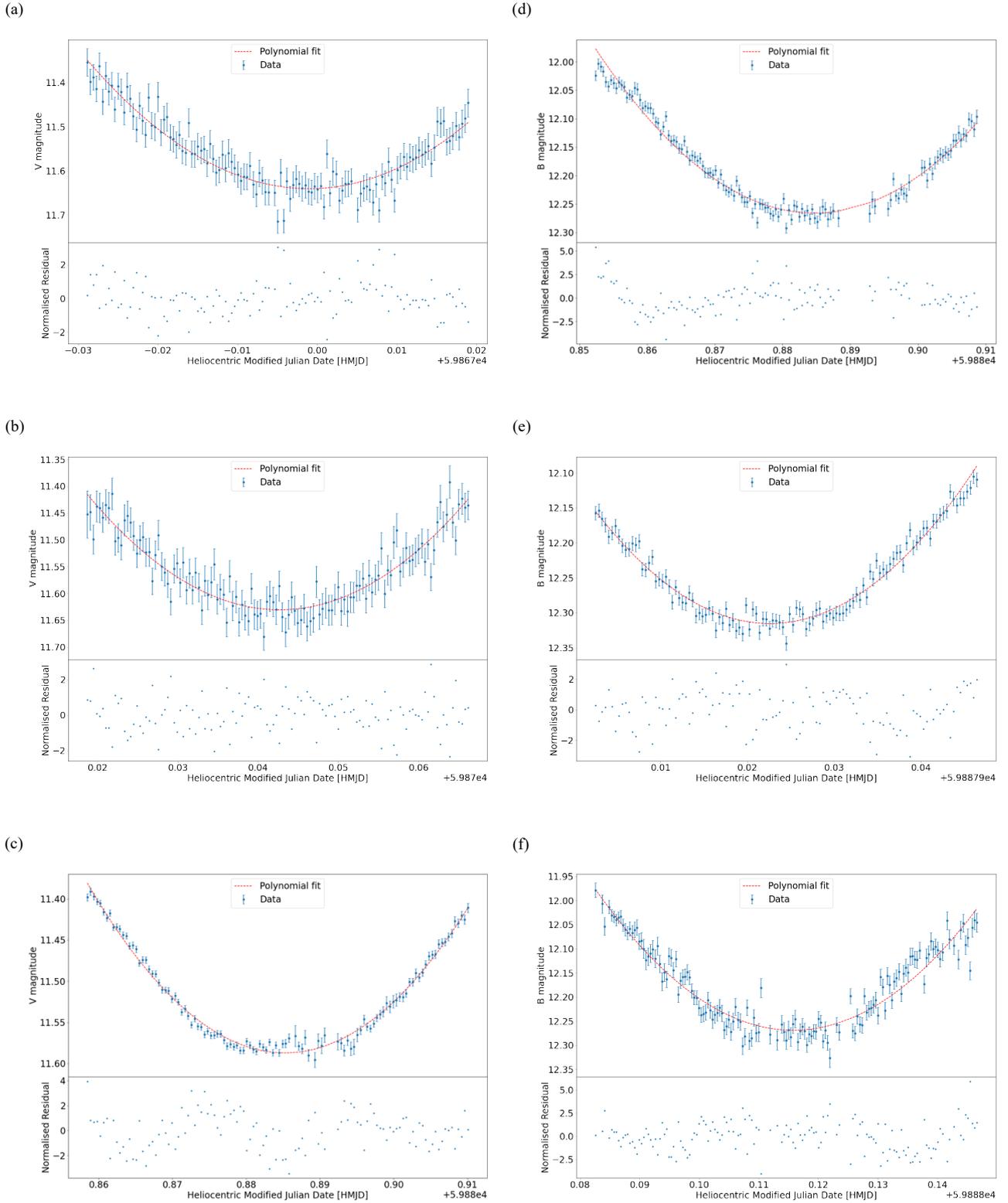

**Fig. 12.** The eclipsing timing measurements of LO And observed: (a) in V-filter over the night of October 14[th] 2022 using Draco 2, (b) in V-filter over the night of October 17[th] 2022 using Draco 2, (c) in V-filter over the night of October 28[th] 2022 using East-16, (d) in B-filter over the night of October 28[th] 2022 using Draco 2, (e) and (f) in B-filter over the night of November 4[th] 2022 using Draco 2.

**Lay Summary for a General Audience**

Binary stars are special celestial objects where two stars orbit around each other in their systems. Among all types of binary stars, only a few of them, such as the visual binaries, can be observed directly to the naked eye, although they are always viewed as single objects in the sky which need professional telescopes to distinguish them from other stars. Most of the binary stars cannot be observed by direct methods, and hence other indirect techniques must be applied to verify their existence. The special type of binary stars this study focused on is the eclipsing binaries, which can be detected when the orbital plane of binary stars happens to be parallel or nearly parallel to our line of sight so that we can measure their magnitude changes during the transits and occultations of each eclipsing component. The magnitude changes of binary stars can be reflected in the variations of their observed light curves, and that's why using necessary photometric analysis to obtain the accurate light curves of eclipsing binaries is the research basis in this study. After this, by performing specific model functions to fit the varying trends of observed light curves, fundamental stellar parameters can be evaluated to fully investigate the physical properties of eclipsing binaries and extrapolate the stellar evolution of other binary stars.